\documentclass[aps,prl,twocolumn,showpacs,floatfix,preprintnumbers,amsmath,amssymb,superscriptaddress,reprint]{revtex4-1}
\usepackage{amsmath}
\usepackage{amssymb}
\usepackage{graphicx}
\usepackage{epstopdf}
\usepackage{epsfig}
\usepackage{dcolumn}
\usepackage{bm}
\usepackage{amsfonts}
\usepackage{mathrsfs}
\usepackage{enumerate}
\usepackage{amsthm}

\begin{document}
\title{Barrier induced chaos in kicked rotor : classical subdiffusion and quantum localization}

\author{Sanku Paul}
\email{sankup005@gmail.com}
\affiliation{Indian Institute of Science Education and Research, Dr. Homi Bhabha Road, Pune 411 008, India}

\author{Harinder Pal}
\email{harinder101@gmail.com}
\affiliation{Instituto de Ciencias F\'isicas, Universidad Nacional Aut\'onoma de M\'exico, CP 62210, Cuernavaca, M\'exico}

\author{M. S. Santhanam}
\email{santh@iiserpune.ac.in}
\affiliation{Indian Institute of Science Education and Research, Dr. Homi Bhabha Road, Pune 411 008, India}

\date{\today}
\begin{abstract}
The relation between classically chaotic dynamics and quantum localization is studied
in a system that violates the assumptions of Kolmogorov-Arnold-Moser (KAM) theorem, namely,
kicked rotor in a discontinuous potential barrier. We show that
the discontinuous barrier induces chaos and more than two distinct subdiffusive energy
growth regimes, the latter being an unusual feature for Hamiltonian chaos.
We show that the dynamical localization in the quantized version of this system
carries the imprint of non-KAM classical dynamics through the dependence of quantum
break time on subdiffusion exponents. We briefly comment on the experimental feasibility
of this system.

\end{abstract}

\maketitle

The interplay between disorder, in the form of chaotic classical dynamics, and
quantum localization continues to attract attention due to a rich
variety of its manifestations. Kicked rotor (KR), in which a particle is periodically
kicked by an external field, is a paradigmatic example for both chaos and 
localization \cite{kr}. From dynamics point of view, the classical KR, for
large kick strengths, displays chaos and energy diffusion \cite{reichl}.
The suppression of diffusion in the quantum regime results from destructive
quantum interference and is termed dynamical localization (DL) due to its analogy with
Anderson localization \cite{reichl,fishman}. Dynamical localization was
experimentally observed in an atom-optics based realization of KR \cite{raizen}.
The emergence of
quantum localization in the variants of KR has led to novel scenarios for
quantum ratchets \cite{tsm}, classical-quantum correspondence \cite{cqc},
coherent quantum control \cite{qcon}, metal-insulator transition \cite{mit},
nonlinearity effects \cite{nlin_eff} and decoherence \cite{qent}. Recently,
an unusual classical 'dynamical localization' was
reported as well \cite{cdloc}.

Kicked rotor is a system that obeys Kolmogorov-Arnold-Moser (KAM)
theorem \cite{reichl} implying that the transition from regular to predominantly chaotic dynamics happens
gradually by breakup of invariant tori upon variation of a control parameter.
We have obtained a good understanding of diffusion
and localization effects in KR as a representative KAM system \cite{reichl}.
Comparatively, much less is known about chaotic systems that violates the assumptions of KAM theorem, the
so-called non-KAM systems. Theoretical studies of non-KAM systems such as the kicked
oscillator \cite{kho}, kicked particle in potential well configurations \cite{sankar,hpal}
have revealed an abrupt transition from integrability to chaos leading to global transport due
to the absence of invariant tori that fragments phase space. In fact, experiments on
non-KAM systems have exploited this property to enhance or in general control electronic transport in
semiconductor superlattices \cite{fromhold} and in coupled billiards in the form of
two-dimensional electron gas in an external magnetic field \cite{brunner}.

Inspite of global classical transport, quantum localization
can suppress transport. In chaotic systems, localization takes various forms.
In time-dependent systems such as KR, DL is a purely quantum
effect that disregards classical dynamics beyond certain time-scale called break-time. 
In autonomous chaotic systems like
the atoms in strong magnetic fields and coupled oscillators \cite{qosc}
semi-classical scarring localization arises due to the influence of isolated
unstable periodic orbits \cite{scar}. Partial barriers in classical phase space, cantori \cite{cantori},
can also lead to localization as in the case of ionic motion in a Paul trap \cite{cantori1}
and in a special case of Bunimovich billiards modelled as a discontinuous
quantum map \cite{dqm}. The classical analogue of the latter system violates KAM theorem.

Generally, quantum-classical correspondence in non-KAM systems has not been studied in detail
and promises new insights in view of the rich classical dynamical features.
Motivated by this, we report on a new non-KAM classical feature, subdiffusive
transport induced by discontinuous potential barriers rather than by cantori, and its relation to
quantum localization. We note that in Hamiltonian systems anomalous transport
arising due to the presence of sticky islands in chaotic sea
is generally superdiffusive in nature and subdiffusive behavior is not seen \cite{altmann}.

Currently, quantum localization in disordered media
with correlated disorder is vigorously investigated, both in theory and experiments \cite{izr2}.
Thus, our results are also relevant in
the broader context of the continuing interest in anomalous diffusion \cite{andiff} and 
localization properties of wave phenomena in different areas \cite{locprop}.

\begin{figure}
\includegraphics*[width=2.0in]{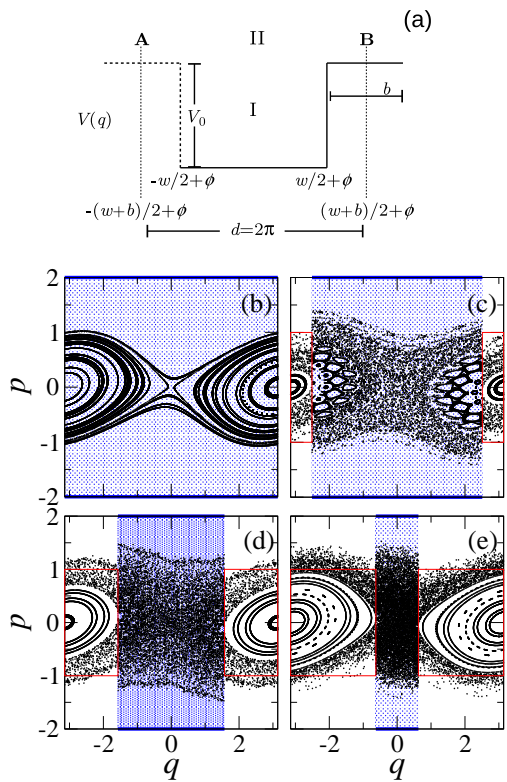}
\includegraphics*[width=2.6in]{fig1b.eps}
\caption{(Color online) (a) Schematic of the stationary
potential, $V(q)$ with $\phi=0$ and $\lambda=2\pi$. Periodic boundary conditions imposed
at positions $A$ and $B$. The regions below and above $V_0$ are
denoted by I and II. (b-e) display the stroboscopic sections.
The barrier regions are marked as (red) rectangular boxes
and the well region is hatched (blue). The parameters
are $V_0=0.5, \phi=0.0$ and $\epsilon=0.25$. The width of well region $w$ is (b) $2\pi$,
(c) $1.6\pi$, (d) $\pi$ and (e) $0.4\pi$.}
\label{fig1}
\end{figure}

The dimensionless Hamiltonian for kicked rotor in potential barriers is
\begin{equation}
H =  \frac{p^2}{2} + V(q) + \epsilon \cos\left(q \right)  \sum_{n=-\infty}^{\infty} \delta(t-n),
\label{skkp}
\end{equation}
where $H_0 = \frac{{p}^2}{2} + V(q)$ is the unperturbed system with
\begin{eqnarray}
V(q) = V_0 \left[ \theta(q-R\pi-\phi) - \theta(q-R\pi-b-\phi) \right]
\label{V2}
\end{eqnarray}
being the stationary potential depicted in Fig. \ref{fig1}(a), $\theta(.)$ is the unit
step function, $V_0$ and $b$ are the
height and width of the potential barrier respectively, $\epsilon$ is
the kick strength, $\phi$ is the phase of the kicking field and $R=w/\lambda$ is the
ratio of width of the well to the wavelength of the kicking field. Physically, Eq. \ref{skkp}
represents a kicked particle in a potential $V(q)$ with periodic boundary conditions applied
at positions $q=\pm (w+b)/2$. Throughout this paper, we choose $\phi=0, w+b=2\pi, \lambda=2\pi$
and consequently $0 < R < 1$.
For convenience, we denote the regions below and above the barrier height $V_0$ as I and II.
In general, the dynamics of the system is determined by $V_0$, $\epsilon$,
and either $R$ or $b$. The classical dynamics can be explicitly reduced to a
map on a suitably chosen stroboscopic section.

If $V_0=0$, then Eq. \ref{skkp} reduces to a kicked rotor on an infinite cylinder.
For $\epsilon=0$, it is classically integrable system.
If $V_0>0$, the potential $V(q)$ is non-analytic and violates the assumptions of KAM theorem.
Thus, when external kicks are introduced with $\epsilon > 0$, KAM tori that
partition the phase space are destroyed, even if $\epsilon$ is arbitrarily small, leading to
chaotic dynamics.
Figure \ref{fig1}(b-e) shows stroboscopic sections obtained by evolving the classical map
for kick strength $\epsilon=0.25$ with several values of $w$. The nature of dynamics can
be understood in terms of that of kicked rotor ($V_0=0$ in Eq. \ref{skkp}).
We recall that for identical value of kick strength, and indeed for any value of $\epsilon << 1$,
the phase space of kicked rotor is covered with invariant tori $I{\!_{KR}}(\omega)$ characterized
by winding number $\omega$. 
In the presence of barriers, a particle initially on a kicked rotor tori $I{\!_{KR}}(\omega)$
would continue to evolve on it until interrupted by the barrier discontinuities at $q=-w/2$ or $w/2$.
This results in reflection
$(q \to q, p \to -p)$ or refraction $(q \to q, p \to \pm\sqrt{p^2-2V_0})$ of the particle
and it hops onto another tori $I^{'}{\!_{KR}}(\omega')$, with $\omega \ne \omega'$.
Every barrier encounter leads to tori hopping. As $n >> 1$, multiple barrier encounters
and the resulting tori hopping ensure that the
autocorrelations decay quickly, resulting in chaotic dynamics (Fig. \ref{fig1}(c-e)) and
diffusion of the energy absorbed from periodic kicks.
This is one manifestation of non-KAM type dynamics in which
barriers play a crucial role in genesis of chaos and energy diffusion.

\begin{figure}
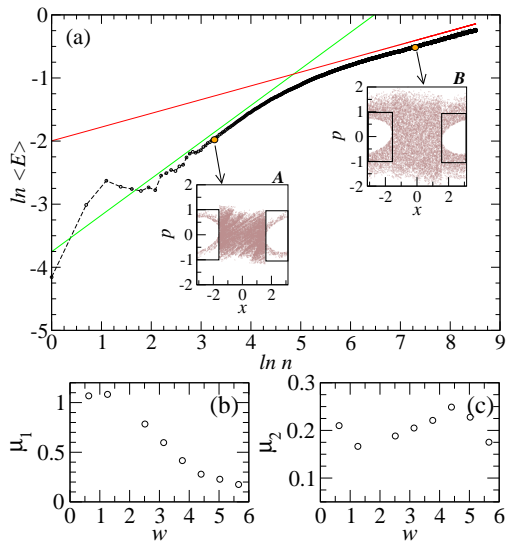

\includegraphics*[width=2.6in]{fig2a.eps}
\includegraphics*[width=2.6in]{fig2b.eps}
\caption{(Color online) The growth of mean energy $\langle E \rangle$ vs. $n$ for parameters
$V_0=0.5, w=\pi$ and $\epsilon=0.25$.
The two solid lines (green and red) fit the linear regimes in log-log plot.
The insets display the stroboscopic sections at 25th kick (A)
and at 1500th kick (B). (b,c) The dependence of $\mu_1$ and  $\mu_2$ on $w$.}
\label{fig2}
\end{figure}

However, if the condition $\pm R\pi + \phi = \pi l$,
$l \in \mathbb{Z}$ is satisfied in region I, then
$I_{KR}(\omega)$ are preserved even with the barrier encounters \cite{hpal}.
This leads to KAM-like tori in region I as shown for $b=0.0$ in Fig. \ref{fig1}(b).
It can be shown that the dynamics is completely hyperbolic
if $R < 0.5$ and $\phi=0$ \cite{sankar}. Physically interesting scenario for energy transport arises if
$l \notin \mathbb{Z}$ leading to non-KAM chaos in region I, and region II displays
invariant tori but punctured by the discontinuities in the potential. 
In the rest of the paper, we choose parameters (listed in Fig. \ref{fig2}) satisfying these conditions.
This choice ensures absence of complete dynamical barriers and cantori
are not effective as barriers to transport. Under these conditions, as kicks impart energy,
global subdiffusive transport becomes possible since the chaotic particles in region I 
penetrate through the punctured tori into region II.

\begin{figure}
\includegraphics*[width=1.3in,angle=-90]{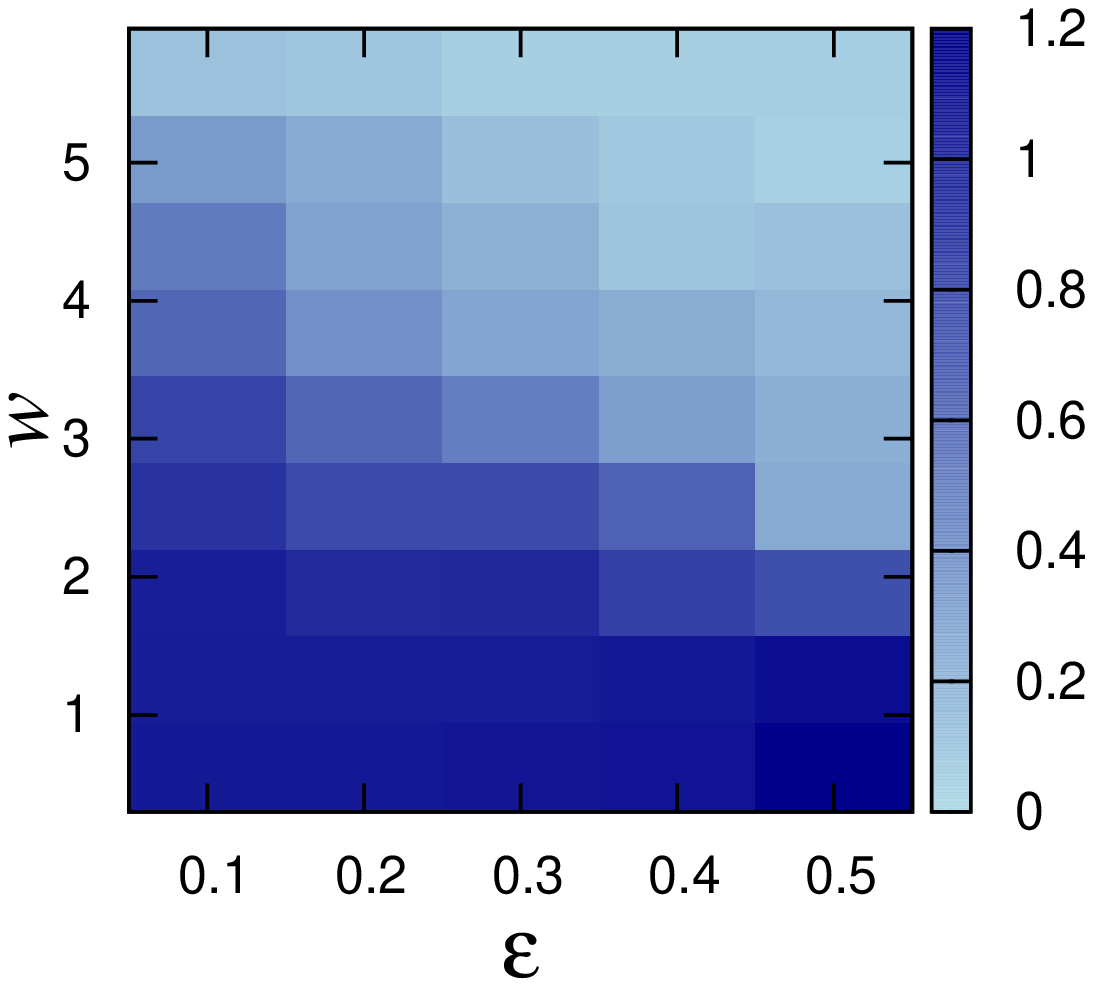} \nolinebreak
\includegraphics*[width=1.3in,angle=-90]{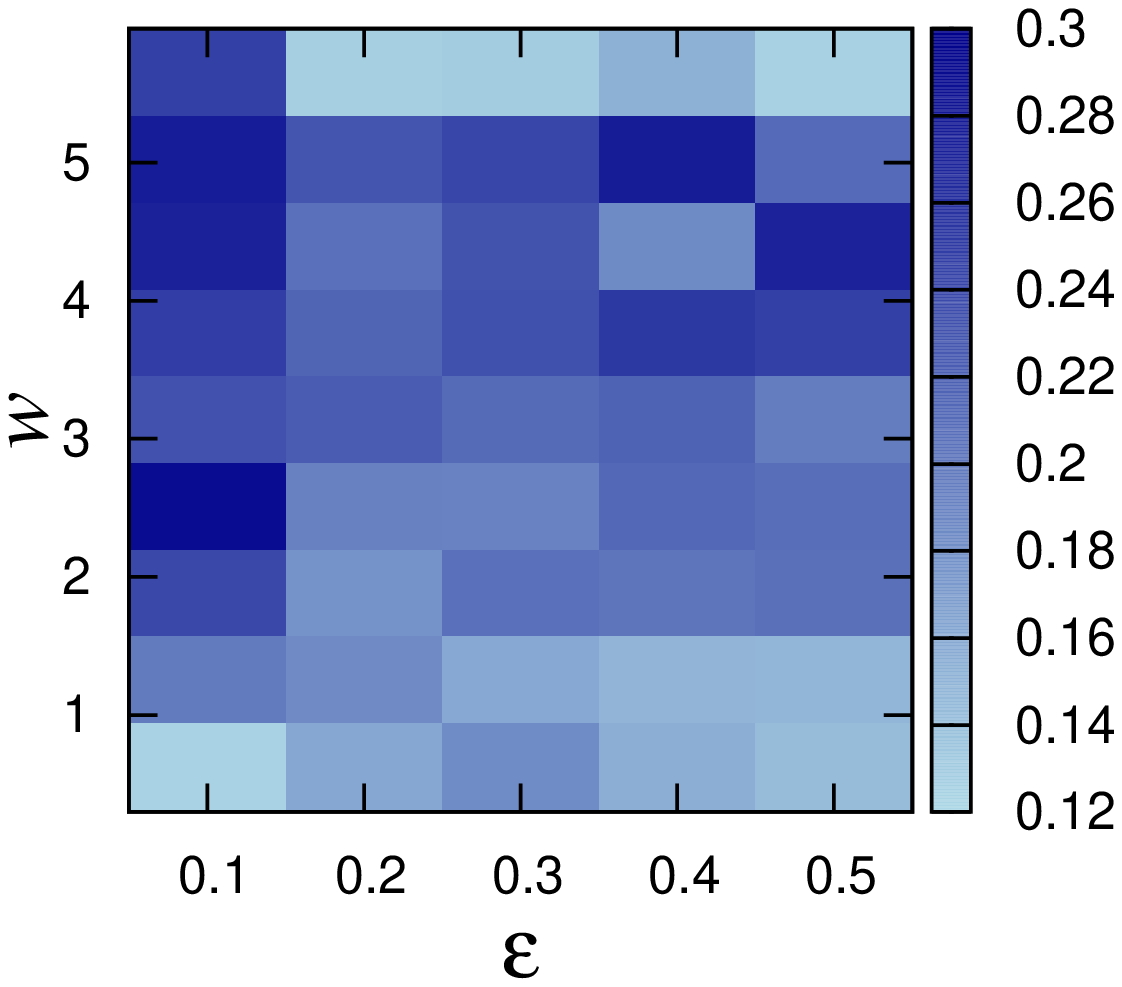}
\caption{(Color online) Subdiffusion exponents $\mu_1$ (left) and $\mu_2$ (right)
plotted as a function of $w$ and $\epsilon$.}
\label{mu12}
\end{figure}

The classical mean energy growth is sub-diffusive and displays two distinct power-law 
regimes within experimentally accessible timescales.
Fig. \ref{fig2}(a) illustrates classical mean energy $\langle E \rangle$ as a function
of time (in units of kick period) for the same set of parameters as in Fig. \ref{fig1}(d).
The initial condition
is an ensemble of points with $-w/2 < q < w/2$ and $p=0$. The first sub-diffusive regime
in Fig. \ref{fig2}(a) and the accompanying stroboscopic section (inset A) show
that most of the particles are physically confined to the well in region I.
Note that the mean energy growth can be described by
$\langle E \rangle_n \sim D_1 ~n^{\mu_1}$, where $D_1$ is the diffusion coefficient
and $0 < \mu_1 \le 1$ is the exponent. 
This subdiffusive behavior can be attributed to correlations.
In general, the phase space in region I can be mixed (intricate
chain of islands in a chaotic sea as in Fig. \ref{fig1}(c)) for $R>0.5$ or
completely chaotic for $R<0.5$ if $V_0 >> 1$.
When $R>0.5$, the energy growth is suppressed significantly due to the correlations induced by a combination of factors, $(i)$ dynamics on $I_{\!_{KR}}(\omega)$ between successive tori hops, 
 $(ii)$ stickiness in the vicinity of chain of islands and $(iii)$ slow diffusion through the punctured tori
in region II. When $R<0.5$, even though the phase space appears
chaotic due to tori hops, correlations of type $\langle q_n q_{n+N} \rangle \propto N^{-\gamma}$
exist because the evolution between two successive barrier
encounters is confined to a kicked rotor tori $I_{\!_{KR}}(\omega)$.
By tuning the number of barrier encounters in one kick period
we can enhance or suppress correlations. As $w\to 0$, the barrier encounters 
and tori hops in one kick period increase and hence
$\mu_1\to 1$ (see Figs \ref{fig2}(b) and \ref{mu12}(a)), the quasi-linear
diffusion limit \cite{kr,reichl} expected under conditions of predominant chaos.

We formally define $\tau_1 = (D_1/D_2)^{1/(\mu_2-\mu_1)}$ as the time
at which first sub-diffusive regime characterized by
$\{D_1, \mu_1\}$ crosses over to the second characterized by $ \{D_2,\mu_2\} $.
The second regime begins after the particles
enter region II for $n > \tau_1$, as seen in the inset B in Fig. \ref{fig2}.
The energy growth is subdiffusive with $\langle E \rangle_n \sim D_2 ~n^{\mu_2}$,
$0 < \mu_2 <1$. In region II, we expect the phase space to display kicked rotor 
tori $I_{\!_{KR}}(\omega)$ punctured by the discontinuities in $V(q)$.
Note that one complete crossing of a barrier of width $b>0$ involves two
refractions, at say, $q=w/2$ and $q=(w/2)+b$. When a particle with energy $E_0$ enters
the above-barrier region and assuming that it does not suffer any kick while transiting
this region, the net change in position between the last and the next kick is denoted by $\Delta q$.
If $\Delta q_{\!_{KR}}$ represents
a similar quantity for the kicked rotor ($V_0=0$), we can show that
$\delta q = | \Delta q - \Delta q_{\!_{KR}}| = b (\beta-1)$, where $\beta=(1-V_0/E_0)^{-1/2}$.
Clearly, if $E_0 \gtrsim V_0$, then $\delta q >> 0$ and if $E_0 >> V_0$ we have $\delta q \to 0$.
This difference leads to torus hopping. Further, as $\delta q \ne 0$, it translates into
momentum difference $\delta p = | \Delta p - \Delta p_{\!_{KR}}| = 
\epsilon | \sin(q_k - (p/|p|) \delta q) - \sin q_k | $
at the position of next kick $q_k$.
Thus, if $E_0 \gtrsim V_0$, the invariant tori do not survive
due to large tori hopping induced by the discontinuities. This appears as chaotic
dynamics in region II in inset B of Fig. \ref{fig1}.
On the other hand, if $E_0 >> V_0$, then invariant tori are
largely preserved with minor dispersion
and hence the tori hopping are a neglible effect. Similar result can be obtained even
if the particle suffers kicks while transiting the barrier region.
For $n > \tau_1$, ignoring a short transient, multiple barrier induced refractions interspersed 
with dynamics on $I_{\!_{KR}}(\omega)$ leads to classical
energy growth that is slower than normal diffusive growth. This regime lasts until a time
scale of $\tau_2 \approx 10^4 \tau_1$ (in units of kick period) by which time the
particles evolve to the vicinity of resonance structure at $|p|=2\pi$.
This can be thought of as long-lasting since $\tau_2$
is much longer than experimentally relevant time scale with atom optics as the
test bed. This subdiffusion mechanism has a weak though systematic
dependence on $w$. As Figs. \ref{fig2}(b) and \ref{mu12}(b) reveal,
$\mu_2 \approx 0.2$ to a first approximation.

To the best of our knowledge, long-lasting subdiffusion in chaotic Hamiltonian systems has
not been reported before, though it was observed in the quantum dynamics of 
nonlinear disordered systems \cite{mario}. Generally, anomalous diffusion in chaotic
Hamiltonians are of superdiffusive type due to
stickiness or the presence of accelerator modes in phase space \cite{altmann,amode,cantori}. 
Further, for bounded and ergodic
systems, superdiffusion follows as a consequence of Kac's theorem \cite{cantori,kac} which guarantees
that the mean recurrence time
exists. However, the phase space of the system in Eq. \ref{skkp} being an infinite cylinder,
is not bounded and Kac's theorem does not strictly apply.
The subdiffusion induced by the potential barriers combined with unbounded
phase space is associated with diverging mean recurrence times (not shown here).


Next, we focus on the quantum regime of the Hamiltonian in Eq. \ref{skkp}.
The period-1 Floquet operator for this kicked system can be obtained as
\begin{equation}
\widehat{U} = \exp\left(-\frac{i\epsilon}{\hbar_s} \cos \widehat{q} \right)
        \exp\left( -\frac{i}{\hbar_s} \left[ \frac{\widehat{p}^2}{2} + \widehat{V} \right] \right),
\label{floquet}
\end{equation}
where $\hbar_s=\frac{2\pi^2 \hbar}{E_c T}$ is the scaled Planck's constant and $E_c=m\lambda^2/2T^2$ 
($T$ is the kicking period).
In this, $\psi(q,n) = \hat{U}^n \psi(q,0)$ for any arbitrary initial wave-packet $\psi(q,0)$.
The classical limit will correspond to taking $\hbar_s \to 0$ keeping $\epsilon$ constant.
Firstly, we solve
the Schr\"odinger equation for the unperturbed system, $H_0 u_m = \lambda_m u_m$ using
momentum eigenstates as the basis, i.e, $u_m(q) = (1/\sqrt{2\pi}) \sum_p a_{m,p} ~ e^{-ipq}, 
p=0,\pm 1,\pm 2,\dots$, where $a_{m,p}$ are the expansion coefficients.
The Floquet operator written in the basis of $u_m$ is
\begin{equation}
U_{mn} = e^{-i\lambda_n/\hbar_s} \sum_{p,p'} ~ a^{*}_{m,p} a_{n,p'} (-i)^{|p-p'|} ~
     J_{|p-p'|}\left(\frac{\epsilon}{\hbar_s}\right)
\label{floquet_matele}
\end{equation}
in which $J_l(.)$ is the Bessel function of order $l$. By numerically solving the
eigenvalue equation $\widehat{U} \xi_i  = e^{i\phi_i} \xi_i$, we obtain the Floquet phases $\phi_i$
and Floquet states $\xi_i$ for $i=1,2,3...N$, where $N$ is the number of basis
states (eigenstates of $H_0$).

\begin{figure}
\includegraphics*[width=3.0in]{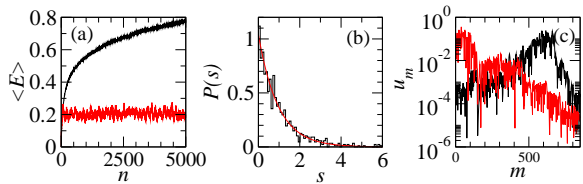}
\caption{(Color online) (a) The classical (black) and quantum (red) mean energy growth
with parameters $V_0=0.5, \phi=0.0$, $w=\pi$, $\epsilon=0.25$ and $\hbar_s=0.0067$.
(b) The nearest neighbor spacing distribution of floquet phases (histogram) and Poisson distribution (solid line).
(c) Two Floquet states in the unperturbed basis (in semi-log scale).
}
\label{fig4}
\end{figure}

We evolve an initial wave-packet $\psi(q,0)$, chosen to be the ground state of $H_0$, under
the action of $\widehat{U}$ and its quantum mean energy $\langle E \rangle$ is displayed
in Fig. \ref{fig4}(a). Clearly, the quantum $\langle E \rangle$ does not follow the 
sub-diffusive behavior of the classical dynamics
beyond certain time scale and instead saturates indicating a localization effect
in energy basis.
From the point of view of random matrix theory, this system
falls in the class of circular orthogonal ensemble (COE). Though we expect
Wigner distribution for the spacings of Floquet phases, quantum localization
of the Floquet states leads to uncorrelated spacings and 
Poisson distribution. This is shown in Fig. \ref{fig4}(b) and
is similar to that of kicked rotor with connections to Anderson
localization \cite{fishman}. The floquet states in the unperturbed basis,
shown in Fig. \ref{fig4}(c), display exponential localization.
As $\hbar_s \to 0$, all the floquet states are localized though the localization lengths
diverge as $\hbar_s^{-2/(2-\mu)}, \mu$ being one of the subdiffusion exponents, and the 
spectral statistics transits from Poisson to COE distribution.

\begin{figure}
\includegraphics*[width=2.3in,]{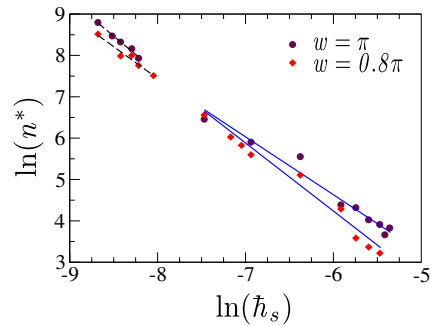}
\caption{(Color online) Scaling of quantum break-time with the scaled Planck's constant $\hbar_s$
for two well widths $w$. The parameters are $V_0=0.5, \phi=0.0$ and $\epsilon=0.25$.
The solid lines and dashed lines have slope $-2/(2-\mu)$ (see Eq. \ref{brtime}), with $\mu=\mu_1$
and $\mu=\mu_2$ respectively (See text for details).}
\label{hbarscale}
\end{figure}

We discuss how non-KAM nature of Hamiltonian in Eq. \ref{skkp} manifests in the quantum domain.
As shown before, the classical subdiffusion of energy arises due to
hopping between KR tori $I_{\!_{KR}}(\omega)$ or punctured tori, a feature facilitated by
the discontinuous potential barriers.
The time scale over which the quantum dynamics follows classical is related
to the inverse of mean spacing of the Floquet spectrum. Assuming that the
mean level density $\rho(E)$
of the unperturbed system near the ground state is proportional to $E^{-1/2}$, we obtain an estimate
for quantum break-time $n^{*}$ to be
\begin{equation}
n^{*} \sim \sqrt{\alpha} \left( \frac{D_i}{2 \pi^2 \hbar_s^2} \right)^{\frac{1}{2-\mu_i}},
\label{brtime}
\end{equation}
where $i=1$ or 2 corresponding to any one of the two classical sub-diffusive regimes and $\alpha$ is a constant.
Significantly, the non-KAM nature of the system leaves its imprint in the quantum domain
through the dependence of $n^{*}$ on diffusion exponents $\mu_1$ or $\mu_2$. For relatively 
large values of $\hbar_s$,
$n^{*} < \tau_1$ and the scaling exponent is $-2/(2-\mu_1)$ depending on the first classical
sub-diffusive regime with exponent $\mu_1$. As $\hbar_s \to 0$, we have $n^{*} > \tau_1$ and
the scaling exponent is $-2/(2-\mu_2)$, corresponding to the second classical sub-diffusive regime
with exponent $\mu_2$.
As shown in Fig. \ref{hbarscale}, a log-log plot of $n^{*}$ against $\hbar_s$ shows linear behavior and agrees well with
the theoretically expected slope $-2/(2-\mu)$, $\mu=\mu_1$ or $\mu_2$. 
This is shown for two sets of parameters. Thus, by varying either $w$ or
$\lambda$, the classical diffusion rate can be controlled leading to a tunable
quantum break-time $n^{*}$.
As $\hbar_s \to 0$, the quantum energy diffusion rate
and the length of diffusive time scale $n^{*}$ can both be controlled by manipulating
system parameters.

In the limit of $\epsilon >> 1$, the system displays predominantly
chaotic dynamics. The potential barriers play only a marginal role in the
genesis of chaos and we obtain a single normal diffusion regime similar to the
case of the kicked rotor.
As would be expected, the quantum dynamics displays dynamical localization.
The central results discussed here, barrier induced chaos and subdiffusive dynamics and its
quantum manifestations, would be valid 
for a larger class of kicked particles placed in barrier-type potentials of various configurations.
We comment on the experimental feasibility of this system. The
kicked rotor was realized using cold atoms in optical lattices created by
two counter-propagating pulsed laser beams \cite{raizen}.
To experimentally realize kicked particle in potential barriers, an additional set of laser beams, that
are always on, can be used to create confining potential with large barrier height.
Periodically kicked particle in a single confining potential barrier was experimentally
realized nearly a decade back \cite{raizen1}. While this set-up could be extended, it is also
possible to achieve this using semiconductor heterostructures \cite{fromhold}.

In summary, we study the classical and quantum dynamics of a
kicked particle interacting with a discontinuous potential.
In this non-KAM system, chaos and subdiffusion are induced by the encounters of the particle
with the discontinuous barriers. The classical mean energy displays more than two distinct
regimes of {\it subdiffusive} growth, a unique feature not seen in Hamiltonian 
systems before. In the quantum domain, the floquet states are
localized in the energy basis. Significantly, the non-KAM nature of the
system manifests in the quantum regime through the dependence of quantum
break-time on subdiffusion exponents, leading to a tunable break-time. These results
can be generalized for various discontinuous potential configurations.

\acknowledgments
HP acknowledges UNAM/DGAPA/PAPIIT research grant number IG100616 and post-doctoral fellowship from
DGAPA-UNAM. SP would like to thank CSIR-UGC for the research fellowship.

\end{document}